\documentclass[preprint]{aastex}
\usepackage{mkfig}

\begin{document}

\title{\bf Heliospheric, Astrospheric, and Insterstellar Lyman-$\alpha$
  Absorption Toward 36~Oph\altaffilmark{1}}

\author{Brian E. Wood, Jeffrey L. Linsky}
\affil{JILA, University of Colorado, and NIST, Boulder, CO 80309-0440.}
\email{woodb@marmot.colorado.edu, jlinsky@jila.colorado.edu}

\and

\author{Gary P. Zank}
\affil{Bartol Research Institute, University of Delaware, Newark, DE 19716.}
\email{zank@bartol.udel.edu}

\altaffiltext{1}{Based on observations with the NASA/ESA Hubble Space
  Telescope, obtained at the Space Telescope Science Institute, which is
  operated by the Association of Universities for Research in Astronomy,
  Inc., under NASA contract NAS5-26555.}

\begin{abstract}

     We use high-resolution UV spectra taken by the Space Telescope Imaging
Spectrograph instrument on board the {\em Hubble Space Telescope} to study
the 5.5 pc line of sight to the K0 V star 36~Oph~A.  The one interstellar
component detected for this line of sight has a velocity inconsistent with
the local interstellar cloud (LIC) flow vector, but consistent with the flow
vector of the so-called G cloud, a very nearby warm cloud in the Galactic
Center direction.  From our data, we measure the following values for the
interstellar temperature, nonthermal velocity, H~I column density,
and D/H value:  $T=5900\pm 500$ K, $\xi=2.2\pm 0.2$ km~s$^{-1}$,
$\log N_{\rm H}=17.85\pm 0.15$, and ${\rm D/H}=(1.5\pm 0.5)\times 10^{-5}$.
The temperature of the G cloud is somewhat lower than that of the LIC, and Mg
and Fe depletions are also lower, but the D/H value appears to be the same.
Based on upper limits for the LIC absorption, we estimate the distance to the
edge of the LIC to be $d_{edge}<0.19$ pc, which the Sun will reach in
$t_{edge}<7400$ yrs.

     The H~I Lyman-$\alpha$ absorption line has properties inconsistent with
those of the other absorption lines, indicating the presence of one or more
absorption components not seen in the other lines.  We present evidence that
hot neutral hydrogen local to both the Sun and the star is responsible for
the excess Lyman-$\alpha$ absorption.  This hot H~I is created by the
interaction between the ISM and the winds of the Sun and 36~Oph~A.  The
observed line of sight lies only $12^{\circ}$ from the upwind direction of
the LIC flow vector, where hydrodynamic models of the heliosphere suggest
that heliospheric H~I absorption should be particularly prominent.  The
properties of the heliospheric absorption ($T=38,000\pm 8000$ K and
$\log N_{\rm H}=14.6\pm 0.3$) are consistent with previous measurements of
this absorption for the $\alpha$~Cen line of sight $52^{\circ}$ from the
upwind direction.

\end{abstract}

\keywords{ISM: atoms --- solar wind --- stars: individual (36~Oph~A)
  --- stars: winds, outflows --- ultraviolet: ISM}

\section{Introduction}

     In 1996, high quality spectra of the very nearby star $\alpha$~Cen
taken by the Goddard High Resolution Spectrograph instrument on board the
{\em Hubble Space Telescope} (HST) yielded a serendipitous detection of
neutral hydrogen in the outer heliosphere \citep[][hereafter LW96]{jll96},
thereby providing a unique new way to observationally study the structure and
internal properties of the heliosphere.  Previous studies of heliospheric
neutral hydrogen had relied primarily on Lyman-$\alpha$ backscatter
measurements, which are mostly sensitive to very nearby H~I \citep{jlb85,eq95}.
In the $\alpha$~Cen data, the heliospheric H~I produces an absorption
signature in the stellar Lyman-$\alpha$ line that allows it to be
detected despite being highly blended with absorption from the local
interstellar medium (LISM).  LW96 demonstrated that the properties of the
heliospheric H~I inferred from the data are consistent with the predictions
of heliospheric models \citep{vbb93,vbb95,hlp95,gpz96,gpz99}.  \citet{kgg97}
confirmed this result by directly comparing the data with H~I absorption
predicted by the models, and they also found evidence that H~I in the
``astrosphere'' of $\alpha$~Cen is also accounting for some of the non-LISM
absorption observed in the Lyman-$\alpha$ line.

     The $\alpha$~Cen line of sight lies $52^{\circ}$ from the upwind
direction of the interstellar flow through the heliosphere.  In the
upwind side, the heliospheric H~I absorption is expected to be dominated by
neutral hydrogen in the ``hydrogen wall'' (or ``H-wall'' for short),
located between the heliopause and solar bow shock.  In this region,
interactions with solar wind protons heat and decelerate the plasma component
of the interstellar wind.  Charge exchange with this plasma then heats,
compresses, and decelerates the neutral hydrogen as well.  The high
temperature and decelerated velocity of the H~I in the hydrogen wall both
play important roles in allowing absorption from this material to be
detectable despite being blended with an LISM absorption component with a
much larger column density.  The high temperature broadens the H-wall
absorption substantially, and the deceleration shifts the absorption away
from the LISM absorption, resulting in a noticeable excess of Lyman-$\alpha$
absorption on the red side of the line.

     The deceleration should be at its largest in the upwind direction,
so the Lyman-$\alpha$ signature of the H-wall should become even easier to
detect if one looks closer to the upwind direction than the $\alpha$~Cen
line of sight $52^{\circ}$ away.  Thus, we searched for another target for
HST that would be much closer to the upwind direction.  We chose the
nearby ($d=5.5$ pc) K0 V star 36~Oph~A (=HD 155886), which is only
$12^{\circ}$ from the upwind direction.  Our main goals in observing this
star are to confirm the detection of the solar hydrogen wall claimed in the
analysis of the $\alpha$~Cen data, and to provide additional constraints on
the properties of the H-wall.

\section{Observations and Data Reduction}

     On 1999 October 10, we observed 36~Oph~A with the Space Telescope
Imaging Spectrograph (STIS) instrument on HST.  The STIS instrument is
described in detail by \citet{rak98} and \citet{bew98a}.
We observed the $1160-1357$~\AA\ spectral region with the high-resolution
E140H grating and $0.2^{\prime\prime}\times 0.2^{\prime\prime}$ aperture,
and we used two exposures with the E230H grating and
$0.1^{\prime\prime}\times 0.2^{\prime\prime}$ aperture to observe
the $2430-2943$~\AA\ spectral range.  The former region contains the
Lyman-$\alpha$ line at 1216~\AA\ that is of primary interest to us, while the
latter region contains interstellar lines of Mg~II and
Fe~II.  These metal lines provide important information on
the properties of the LISM that can be used to constrain the LISM H~I
Lyman-$\alpha$ absorption, making it easier to separate it from the
heliospheric absorption.

     The data were reduced using the STIS team's CALSTIS software package
written in IDL \citep{dl99}.  The reduction included assignment of
wavelengths using calibration spectra obtained during the course of the
observations.  A substantial amount of scattered light was clearly present
in the saturated core of the Lyman-$\alpha$ line.  This flux was removed
from the data using the \verb&ECHELLE_SCAT& routine in the CALSTIS package.
A geocoronal Lyman-$\alpha$ emission feature in the middle of the
interstellar H~I absorption was fitted with a Gaussian, and then removed by
subtracting the fitted Gaussian from the data.  The centroid of the Gaussian
is at $-26.6$ km~s$^{-1}$, which agrees very well with the expected location
of $-26.4$ km~s$^{-1}$ based on the Earth's projected velocity toward the
star at the time of observation.  This confirms the accuracy of our wavelength
calibration.

\section{Data Analysis}

\subsection{Single Component Fits}

     Besides H~I Lyman-$\alpha$ at 1215.670~\AA, there are only four
interstellar absorption features apparent in our data that can be accurately
measured:  D~I Lyman-$\alpha$ $\lambda$1215.339, Fe~II $\lambda$2600.173, and
the Mg~II h \& k lines at 2803.531~\AA\ and 2796.352~\AA, respectively.  The
rest wavelengths quoted above are vacuum wavelengths.  These lines are all
shown in Figure 1.

     None of the lines in Figure 1 show any asymmetries that would indicate
the presence of more than one absorption component, so all are fitted with a
single component.  The dotted lines in the figure are the fits
before correction for instrumental broadening, and the thick solid lines
that fit the data are the fits after the instrumental broadening correction
is applied.  The line spread functions used in these corrections were taken
from \citet{kcs99}.  The oscillator absorption strengths assumed in our fits
are from \citet{dcm91}.  The parameters of the fits are listed in Table 1.

     In order to constrain the Mg~II fits as much as possible, the 2 Mg~II
lines were fitted simultaneously with both lines forced to have the same
column density ($N$), Doppler parameter ($b$), and central velocity ($v$).
We were concerned about the possibility that much of the residual flux
present below the Mg~II absorption lines may be scattered light.
Unfortunately, there is currently no scattered light correction for E230H
data like there is for E140H spectra (see above).  We tried fits with much of
the possible scattered light flux subtracted from the spectrum to see the
degree to which the fit parameters were affected.  The uncertainties quoted
in Table 1 include the systematic uncertainties we estimate due to the
uncertain scattered light correction.

     Both LISM studies and {\it in situ} measurements of LISM material within
the solar system suggest that the velocity of the incoming interstellar
wind is about 26 km~s$^{-1}$ \citep{mw93,rl95}.
However, the LISM flow vector appears to be different for lines of
sight toward the Galactic center, with a faster speed of 29.4 km~s$^{-1}$.
Thus, it has been proposed that a different cloud lies in this direction, the
so-called G cloud \citep{rl92}.  If this interpretation is
correct the Sun must be very near the edge of the Local Interstellar Cloud
(LIC), since all the LISM absorption detected by LW96 toward $\alpha$~Cen
(which is roughly in the Galactic center direction) is at the expected
velocity of the G cloud and none is at the velocity predicted for that line
of sight by the LIC vector \citep[see also][]{rl95}.

     The outlines of the LIC and the G cloud in the Galactic plane are shown
in Figure 2, as are the projected locations of the Sun, $\alpha$~Cen, and
36~Oph.  The LIC outline is from \citet{sr00}.  The G cloud shape is not well
known, so the contour shown in Figure 2 is only a crude but plausible
representation for this cloud.  The velocity vectors of the LIC and G clouds
are also shown in the figure.

     With Galactic coordinates of $l=358^{\circ}$, $b=7^{\circ}$; 36~Oph
lies almost directly toward the Galactic center, very near the upwind
directions of both the LIC and G cloud vectors.  The flow velocities
predicted for the 36~Oph line of sight by these two vectors are $-25.1$
and $-28.4$ km~s$^{-1}$, respectively.  The LISM velocity measured toward
36~Oph agrees very well with the latter velocity (see Table 1), so
we conclude that all of the LISM absorption we detect is from the G cloud.
Inspection of Figure 1 reveals that there is no visible absorption component
centered at the LIC velocity.  Thus, our findings are
very similar to those of LW96.  However, the evidence for higher flow
velocities toward the Galactic Center, and the close proximity of the edge
of the LIC, is particularly striking in our data, because even the
{\em projected} LISM velocities that we directly observe are larger than the
26 km~s$^{-1}$ velocity that has been measured for the LIC.

     We now try to estimate how far the Sun is from the edge of the LIC.
We need to first estimate an upper limit for the H~I column density, and the
best way to do this is to determine upper limits for the D~I and/or Mg~II
column densities and then calculate the H~I upper limit based on the known
abundance ratios in the LIC.  For both D~I and Mg~II we repeat the fits
shown in Figure 1, but with added LIC absorption components centered on the
known LIC velocity and with Doppler parameters consistent with previous
observations of LIC material ($b_{\rm D~I}=8.2$ km~s$^{-1}$ and
$b_{\rm Mg~II}=3.0$ km~s$^{-1}$).  We experiment with different LIC/G cloud
column density ratios to see at what point we believe the fits become
unacceptable.

     One might think that the Mg~II lines would provide the best constraints
because they are narrower and more optically thick than D~I.  However,
the D~I/Mg~II ratio is larger in the LIC than in the G cloud by about a
factor of 4 (see below).  Based on experiments with different D~I fits,
we estimate that the D~I column density of the LIC is no more than 10\%
that of the G cloud, corresponding to an upper limit of
$N_{\rm D~I}<9\times 10^{11}$ cm$^{-2}$.  Note that this corresponds to
a LIC contribution of 2.5\% to the Mg~II lines.  For LIC contributions greater
than this, the $\chi^{2}_{\nu}$ values of the fits become significantly worse,
and the G cloud D~I absorption becomes more and more blueshifted away from
the expected velocity of the G cloud.  (If we force it to be at the expected
velocity, the quality of the fit degrades even more.)

     The LIC D/H value is $1.5\times 10^{-5}$ \citep{jll98a,jll98b},
so $N_{\rm H~I}<6\times 10^{16}$ cm$^{-2}$ for the LIC.  Assuming
a density of $n_{\rm H~I}=0.10$ cm$^{-3}$ \citep{jll00},
the upper limit for the distance to the edge of the LIC is
$d_{edge}< 0.19$ pc for the 36~Oph line of sight, consistent with the
$d_{edge}=0.05$ pc value suggested by the LIC model of \citet{sr00} (see
Fig.\ 2).  The LIC material is
moving toward the Sun with a velocity of $-25.1$ km~s$^{-1}$ along this line
of sight, so we will reach the edge in $t_{edge}<7400$ yrs.

     The Doppler parameters ($b$) listed in Table 1 are related to the
temperature, $T$, and nonthermal velocity, $\xi$, by the following equation:
\begin{equation}
b^2 = 0.0165\frac{T}{A} + \xi^2,
\end{equation}
where $b$ and $\xi$ are in units of km~s$^{-1}$, and $A$ is the atomic weight
of the element in question.  In Figure 3, the measured Doppler parameters of
the Mg~II, Fe~II, and D~I lines are used with equation (1) to derive curves
of $\xi$ vs.\ $T$, with error bars.  The shaded region in the figure is the
area of overlap for these curves, which identifies the temperature and $\xi$
values for the observed LISM material.

     Our measured temperature and nonthermal velocity based on this analysis
are $T=5900\pm 500$~K and $\xi=2.2\pm 0.2$ km~s$^{-1}$.
This temperature is consistent with the $T=5400\pm 500$ K temperature
observed toward $\alpha$~Cen, supporting the assertion of LW96 that the G
cloud temperature is somewhat cooler than the LIC temperature of
$T=8000\pm 1000$ K \citep{ard97,np97,bew98}.
However, the nonthermal velocity toward 36~Oph appears to be somewhat
larger than that observed toward $\alpha$~Cen
($\xi=1.25\pm 0.25$ km~s$^{-1}$).

     The relative abundances of D~I, Mg~II, and Fe~II suggested by the column
densities in Table 1 are consistent with the $\alpha$~Cen measurements.
These results suggest that Mg and Fe are less depleted in the G cloud than
in the LIC.  For example, the D~I/Mg~II ratio is $0.9\pm 0.4$ toward
36~Oph and $1.2\pm 0.2$ toward $\alpha$~Cen, but is $\geq 4$ for the
LIC \citep[LW96;][]{ard97,np97}, with the apparent
exception of the Sirius line of sight which has a D~I/Mg~II ratio of about
1.7 \citep{rl94,pb95}.

     In Figure 4, we present our single component fit to the H~I
Lyman-$\alpha$ line, combined with the D~I Lyman-$\alpha$ fit from Figure 1.
The parameters of the H~I fit are provided in Table 1.  In constructing this
fit, we started with an assumed stellar Lyman-$\alpha$ profile based on a
broadened version of the observed Mg~II k line profile.  We then performed an
initial fit, altered the assumed stellar profile to improve the quality of
the fit, and then attempted another fit.  Yet another iteration of this
process was required before arriving at the fit in Figure 4.

     This Lyman-$\alpha$ fitting technique has been used in many past
analyses \citep[LW96;][]{np97,bew98}.  The argument
for using the Mg~II line as a starting point is that the Mg~II k and
Lyman-$\alpha$ lines are both highly optically thick chromospheric lines that
have similar shapes in the solar spectrum.  It should be stated, however,
that for single component fits it actually matters little what initial model
is used for the stellar Lyman-$\alpha$ profile.  The initial profile is
altered significantly before the final best fit is determined, and the final
fit parameters are therefore independent of the initial assumptions about the
shape of the stellar profile.  This was demonstrated clearly in the
$\alpha$~Cen analysis, where the Lyman-$\alpha$ lines of both members of the
$\alpha$~Cen binary system were independently fitted with single absorption
components and the derived interstellar parameters were the same (LW96).

     The residuals of the single component H~I fit in Figure 4 suggest some
minor systematic discrepancies, but the quality of the fit is not too bad.
However, the main problem with the fit is that the velocity and Doppler
parameter of the H~I absorption are completely inconsistent with
those of the other lines, D~I in particular (see Table 1).  The central
velocity of H~I is $-25.9$ km~s$^{-1}$, as opposed to the $-28.4$ km~s$^{-1}$
average velocity observed for the other lines.  The Doppler parameter of
H~I (14.34 km~s$^{-1}$) suggests temperatures of about 12,000 K, compared
with the $\sim 6000$ K temperature determined mostly from D~I (see Fig.\ 3).
If a curve was plotted in Figure 3 for H~I it would be a nearly vertical
line centered at about 12,000 K, which actually lies off the right edge of
the figure, emphasizing just how discrepant H~I is relative to D~I and the
other lines.

     These problems with H~I are very reminiscent of those found for the
$\alpha$~Cen line of sight (LW96).  Collectively they represent the primary
piece of evidence for a hydrogen wall contribution to the Lyman-$\alpha$
absorption toward both $\alpha$~Cen and 36~Oph, because the only way to
resolve the discrepancies is to add a second absorption component to the H~I
fit with properties that turn out to be consistent with those expected for
the solar H-wall.  The H~I Lyman-$\alpha$ line of $\alpha$~Cen exhibits a
$+2.2$ km~s$^{-1}$ velocity discrepancy and about a $+3000$ K temperature
discrepancy relative to the other lines (LW96).  The 36~Oph discrepancies
are in the same direction, but are even larger ($+2.5$ km~s$^{-1}$ and
$+6000$ K).  This is consistent with our expectations, since the H-wall
closer to the upwind direction should be more decelerated and may be a bit
hotter \citep[see, e.g.,][]{gpz96}.

     Since the LIC velocity is redshifted relative to the G cloud velocity,
one might wonder if LIC H~I absorption might be responsible for the velocity
discrepancy between H~I and the other lines.  In Figure 4 we show the LIC
absorption associated with the $N_{\rm H}<6\times 10^{16}$ cm$^{-2}$
upper limit derived above (dotted line).  Essentially all of
the LIC absorption is well within the saturated core of the line, so the
LIC will not have any observable effect on the Lyman-$\alpha$ line.

\subsection{The Bisector Technique}

     Although single component H~I fits are unique and well-constrained,
this is not necessarily the case for two component fits, as LW96 demonstrated
in the $\alpha$~Cen analysis.  Thus, before attempting a two component fit,
we try to constrain the LISM H~I column density by other means.  \citet{bew96}
measured an LISM column density toward
$\epsilon$~Indi by determining the amount of absorption in the wings of the
Lyman-$\alpha$ line, based on the assumption that the far wings of the stellar
Lyman-$\alpha$ profile should be centered on the rest frame of the star, as
is the case for the Sun.  This ``bisector technique'' only works when there
is a substantial wavelength difference between the LISM absorption and the
center of the stellar Lyman-$\alpha$ line, creating a situation where there
is more absorption in one wing than the other.  This induces an apparent line
shift in the wings, the magnitude of which is dependent on the LISM column
density.  This analysis technique could not be tried in the $\alpha$~Cen
analysis, since there is no significant shift between the stellar emission
and LISM absorption, but in our 36~Oph data there is about a 30 km~s$^{-1}$
shift between the two.

     The bisector technique relies on an accurate stellar radial velocity.
The radial velocity listed for 36~Oph in the literature is $-1$ km~s$^{-1}$
\citep{ah91}.  However, all the chromospheric
lines in our spectra, including Mg~II h \& k and the O~I triplet at 1300~\AA,
are centered on $+1.0$ km~s$^{-1}$.  Thus, we believe this is a more likely
value for the radial velocity, and in any case a far more likely value for the
centroid of the Lyman-$\alpha$ wings.  Since 36~Oph~A is a member of a binary
system, we speculate that perhaps orbital motion has changed the stellar
velocity from the systemic value of $-1$ km~s$^{-1}$.  Based primarily on
the centroids of the Cl~I $\lambda$1351.657 and O~I] $\lambda$1355.598 lines,
which can be measured very accurately because of the very narrow line widths
of these features ($FWHM \sim 0.04$ \AA), we assume a velocity of
$+1.0\pm 0.2$ km~s$^{-1}$ for the star.  Since other chromospheric lines in
our spectra have this centroid velocity, including Mg~II h \& k, it is
reasonable to assume the far wings of Lyman-$\alpha$, which are formed at the
base of the chromosphere, will have it too.

     Figure 5 shows how the analysis technique works.  First, we fit
polynomials to the wings of the observed Lyman-$\alpha$ line, interpolating
over the D~I line.  Working from these fits (thick solid lines in Fig.\ 5a),
we can derive what the wings of the stellar Lyman-$\alpha$ line would look
like assuming different values of the H~I column density.  The results for five
different column densities are shown in Figure 5a.  In Figure 5b we display
the bisectors of these wings for various values of $\log N_{\rm H}$.  As
the column density increases the bisectors become more and more blueshifted,
due to the fact that the LISM is absorbing more in the blue wing of the
line than in the red wing.

     Ideally, a vertical bisector centered at the radial velocity of the star
should result when the correct value of $\log N_{\rm H}$ is assumed, but in
practice this does not happen due to uncertainties in the polynomial fits to
the data and in measuring the location of the bisectors.  Nevertheless, in
Figure 5b we identify with a solid line the bisector that we believe best
matches the stellar radial velocity of $+1.0$ km~s$^{-1}$, which corresponds
to $\log N_{\rm H}=17.84$.  Column densities in the range
$\log N_{\rm H}=17.7-18.0$ all produce bisectors with average velocities
within about $\pm 1$ km~s$^{-1}$ of the expected velocity, so we decide on a
value of $\log N_{\rm H}=17.85\pm 0.15$ for our best estimate of the
LISM H~I column density toward 36~Oph.

     This range of columns yields deuterium-to-hydrogen ratios in the range
${\rm D/H}=(1.0-2.0)\times 10^{-5}$, so we quote a value of
${\rm D/H}=(1.5\pm 0.5)\times 10^{-5}$.  This is an encouraging result, since
it is consistent with the mean value ${\rm D/H}=(1.5\pm 0.1)\times 10^{-5}$
for the LIC \citep{jll98a,jll98b}.  The error bars in the G
cloud D/H value are too large to tell if it is actually identical to the LIC
value, but apparently the G cloud does not have a drastically different D/H
value than the LIC.

\subsection{Multi-Component Fits}

     In Figure 6a, we present a two component fit to the Lyman-$\alpha$ line,
where the dotted line represents LISM absorption and the dashed line
represents absorption from the solar hydrogen wall.  The LISM component is
forced to have the average velocity observed for the other LISM lines
($v=-28.4$ km~s$^{-1}$) and a Doppler parameter derived from the T and $\xi$
values measured from the other lines ($b=10.11$ km~s$^{-1}$).  Only a single
two component fit is shown in Figure 6, but in practice we performed
many fits assuming different stellar Lyman-$\alpha$ profiles that allowed the
LISM H~I column density to vary within the $\log N_{\rm H}=17.7-18.0$ range
allowed by the bisector analysis.  This experimentation allowed us to better
determine the best fit parameters and their uncertainties, which are listed
in Table 1.

     The quality of the two component fit is certainly better than the
one component fit in Figure 4, based on both the residuals shown in the
figures and the $\chi_{\nu}^{2}$ values listed in Table 1.  More importantly,
in the two component fit the D~I and H~I lines are self-consistent.  The
H-wall temperature derived from the Doppler parameter of the two component
fit is $T=49,000\pm 9000$ K, compared with $T=29,000\pm 5000$ K toward
$\alpha$~Cen.  It is worth noting, however, that the meaning of these
measured temperatures is not entirely clear.  The absorption component fits
to the Lyman-$\alpha$ line use Voigt functions for the opacity profiles,
which amounts to an implicit assumption that velocity distributions for each
component are Maxwellian, but neutrals in the outer heliosphere are far from
being in equilibrium and therefore do not have Maxwellian distributions
\citep{vbb98,asl98,hrm00}.  Furthermore, the absorption components are
integrated over a long line of sight through the heliosphere, resulting in
even more complex velocity distributions.

     There is a serious problem with the central velocity of the H-wall
absorption inferred from the two-component fit.
Heliospheric models all suggest substantial decelerations within the hydrogen
wall in the upwind direction
\citep*{vbb93,vbb95,hlp95,gpz96,vbb98,asl98,hrm00},
meaning the H-wall absorption should be
significantly redshifted relative to projected velocity of the LIC for that
line of sight, $-25.1$ km~s$^{-1}$.  The measured velocity, $v=-26.7\pm 0.3$
km~s$^{-1}$, is only slightly redshifted relative to the G cloud absorption,
and it is actually {\em blueshifted} relative to the LIC velocity.
Figure 6a illustrates why this is the case.  Non-LISM absorption exists on
both the blue and red sides of the H~I absorption feature, and the
heliospheric component cannot be greatly redshifted away from the LISM
absorption in order to account for the excess absorption on the blue side.

     The only way the heliospheric absorption can have decelerations
consistent with the models is for there to be yet another absorption
component that accounts for the non-LISM absorption on the blue side of
the line.  \citet{kgg97} found this to be the case for the
$\alpha$~Cen data as well when they made direct comparisons between the
data and the H~I absorption predicted by heliospheric models.  \citet{kgg97}
interpreted this to mean that the ``astrosphere'' of $\alpha$~Cen was
responsible for the blueshifted non-LISM H~I absorption.  We propose that
our Lyman-$\alpha$ data are also contaminated by astrospheric absorption.

     Figure 2 shows the velocity vector of 36~Oph, which is computed from
the radial velocity and proper motion information in \citet*{ah91}.
From this 32.5 km~s$^{-1}$ stellar vector and the known 29.4
km~s$^{-1}$ G cloud vector, we can determine the interstellar wind vector
in the rest frame of 36~Oph (dotted arrow in Fig.\ 2).  We find that the wind
speed relative to the star is 40 km~s$^{-1}$ and the line of sight toward the
Sun is $\theta=134^{\circ}$ from the upwind direction, meaning we are
looking at the downwind portion of 36~Oph's astrosphere.

     Much of the heliospheric H~I in downwind directions is formed by charge
exchange between LISM neutrals and solar wind protons inside the heliopause,
whereas the H-wall H~I observed in upwind directions is created by charge
exchange between LISM neutrals and heated LISM protons {\em outside} the
heliopause.  Another difference is that in the upwind direction, the
H~I in the H-wall is decelerated, but models suggest that for downwind
directions the H~I gas is {\em accelerated} relative to the LISM flow.  One
consequence of this is that heliospheric absorption will be redshifted
relative to the LISM absorption in all directions.  Likewise, astrospheric
absorption will always be blueshifted.  \citet{vvi99}
claim to have detected redshifted heliospheric absorption in a
downwind direction toward the star Sirius, consistent with these theoretical
expectations.

     Therefore, it is not unreasonable to suppose that astrospheric material
downwind from 36~Oph is responsible for the non-LISM absorption on the
blue side of the Lyman-$\alpha$ absorption feature.  Thus, in Figure 6b we
present a three component fit to the Lyman-$\alpha$ line, the parameters of
which are listed in Table 1.  Because they are so highly blended, the three
components must be constrained somehow to produce a unique fit.  The LISM
component is constrained the same as in the two component fit.  Furthermore,
we force the heliospheric and astrospheric absorption components to have
velocities roughly consistent with model predictions.

     Heliospheric models suggest an average deceleration of roughly 40\%
within the hydrogen wall in the upwind direction, meaning the projected
velocity toward 36~Oph should be $V_{\rm H}=-0.6V_{0}\cos \theta$, where
$V_{0}=26$ km~s$^{-1}$ is the LISM flow velocity for the Sun and
$\theta=12^{\circ}$ is the angle of the line of sight relative to the upwind
direction.  Thus, $V_{\rm H}=-15$ km~s$^{-1}$ for the heliospheric absorption.
In downwind directions, the heliospheric models suggest accelerations of
about 30\% for the H~I.  Assuming this is the case for 36~Oph, the predicted
velocity of the astrospheric absorption is then
$V_{\rm H}=V_{rad}+(1.3V_{0}\cos \theta)$, where $V_{rad}=1.0$ km~s$^{-1}$
is the stellar radial velocity, $V_{0}=40$ km~s$^{-1}$ is the LISM flow
velocity relative to the star, and $\theta=134^{\circ}$ is the angle relative
to the upwind direction of the solar line of sight.  Thus, $V_{\rm H}=-35$
km~s$^{-1}$ for the astrospheric absorption.  We arbitrarily assume
uncertainties of $\pm 5$ km~s$^{-1}$ for both the estimated heliospheric and
astrospheric velocities.  Analogous to the procedure used in the two
component fit, we perform many fits varying the LISM column density and
the heliospheric/astrospheric velocities within the allowed error bars,
which allows us to determine the best fit parameters and uncertainties
quoted in Table 1.

     Since the heliospheric component no longer has to account for the excess
absorption on both the red and blue sides of the Lyman-$\alpha$ line, the
column density and Doppler parameter are significantly lower than for the two
component model (see Table 1).  The column density
($\log N_{\rm H}=14.6\pm 0.3$) is now consistent with that measured toward
$\alpha$~Cen ($\log N_{\rm H}=14.74\pm 0.24$), as is the temperature inferred
from the Doppler parameter once the quoted uncertainties are considered
($T=38,000\pm 8000$ K for 36 Oph, $T=29,000\pm 5000$ K for $\alpha$~Cen).

     The astrospheric component has a column density
($\log N_{\rm H}=14.7\pm 0.3$) and Doppler parameter ($b=26.0\pm 3.0$
km~s$^{-1}$) very similar to that of the heliospheric component, with the
Doppler parameter implying a temperature of $41,000\pm 10,000$ K.  We would
have expected temperatures somewhat higher than this for a downwind line of
sight, based on the H~I temperatures predicted by the hydrodynamic models
\citep*[e.g.,][]{hrm00}.  However, a direct comparison between
the data and the absorption predicted by the models is needed to see if they
are truly inconsistent.

     The average LISM density toward 36~Oph is $n_{\rm H}=0.03-0.06$
cm$^{-3}$.  This is lower than the $n_{\rm H}\approx 0.1$ cm$^{-3}$ densities
typically observed toward the nearest stars, including $\alpha$~Cen
\citep{jll00}.  One possible explanation for this is that hot
interstellar material without any neutral H occupies part of the line of
sight, either in the foreground of the observed absorption between the LIC
and G clouds; or beyond the G cloud, which would mean that 36~Oph lies outside
the G cloud.  However, if our detection of astrospheric H~I is valid, the
LISM around 36~Oph must necessarily contain H~I, meaning 36~Oph must be
within the G cloud.  Furthermore, the existence of substantial G cloud
material toward $\alpha$~Cen, which is only 1.3 pc away, suggests that a
large gap between the LIC and G clouds toward 36~Oph is unlikely (see
Fig.\ 2).  Thus, the most likely explanation for the low average density
is a negative density gradient within the G cloud toward 36~Oph.
\citet{bew98} used similar reasoning to infer a density gradient within the
LIC toward 40~Eri.

\section{Summary}

     We have analyzed absorption features observed in HST/STIS spectra of
the 5.5 pc line of sight to 36~Oph~A.  Our findings are summarized as
follows:
\begin{description}
\item[1.] Only one LISM absorption component is seen toward 36~Oph, with a
  velocity consistent with the flow vector of the G cloud.  No absorption
  whatsoever is detected from the LIC, meaning the edge of the LIC
  must be very close.  Because the line of sight is
  very close to the upwind directions of both the G and LIC clouds, the
  velocity of $-28.4$ km~s$^{-1}$ that we observe is very close to
  the actual flow speed of the observed material.  The fact that this
  {\em projected} velocity is greater than the 26 km~s$^{-1}$ velocity of the
  LIC provides particularly strong evidence for the existence of faster G
  cloud material in that direction.
\item[2.] We estimate that the LIC D~I column density must be less than
  $9\times 10^{11}$ cm$^{-2}$ to explain why no LIC absorption is detected.
  This corresponds to $N_{\rm H~I}<6\times 10^{16}$ cm$^{-2}$ based
  on the known LIC D/H ratio ($1.5\times 10^{-5}$).  Assuming a density
  of $n_{\rm H~I}=0.1$ cm$^{-3}$, the distance to the edge of the LIC is
  $d_{edge}<0.19$ pc.  The Sun will reach the edge in $t_{edge}<7400$
  yrs, based on the LIC velocity vector.
\item[3.] The temperature ($T=5900\pm 500$ K) and relative abundances of D~I,
  Mg~II, and Fe~II (e.g., ${\rm D~I/Mg~II}=0.9\pm 0.4$) are within the error
  bars of the measurements made for the $\alpha$~Cen line of sight, which
  also samples G cloud material (LW96).  These measurements suggest that the
  G cloud has a slightly lower temperature and less Mg and Fe depletion than
  the LIC.  However, the nonthermal velocity toward 36~Oph,
  $\xi=2.2\pm 0.2$ km~s$^{-1}$, is higher than the $\alpha$~Cen
  measurement ($\xi=1.25\pm 0.25$ km~s$^{-1}$).
\item[4.] Using a Lyman-$\alpha$ bisector technique first described by
  \citet*{bew96}, we estimate a column density of
  $\log N_{\rm H}=17.85\pm 0.15$ toward 36~Oph and a deuterium-to-hydrogen
  ratio for the G cloud of ${\rm D/H}=(1.5\pm 0.5)\times 10^{-5}$.
  The latter quantity is consistent with the LIC value of
  ${\rm D/H}=(1.5\pm 0.1)\times 10^{-5}$.
\item[5.] The average G cloud density toward 36~Oph is $n_{\rm H}=0.03-0.06$,
  significantly less than the $n_{\rm H}\approx 0.1$ cm$^{-3}$ densities
  typically observed toward the nearest stars, including $\alpha$~Cen within
  the G cloud.  The most likely interpretation is that the G cloud H~I
  density decreases toward 36~Oph.
\item[6.] The H~I Lyman-$\alpha$ absorption line is redshifted and suggests a
  much higher temperature than the other lines.  LW96 used a very similar
  discrepancy toward $\alpha$~Cen to argue for the presence of absorption
  from heated, decelerated material in the solar hydrogen wall.  Our results
  strongly support this interpretation.
\item[7.] In two component fits to the H~I Lyman-$\alpha$ line, the solar
  hydrogen wall component is not decelerated relative to the projected LIC
  velocity, which is a serious contradiction with theoretical
  expectations, so we propose that absorption from astrospheric material
  around 36~Oph is responsible for some of the absorption on the blue
  side of the line.  Thus, our best model of the Lyman-$\alpha$ absorption
  has three components:  LISM (G cloud), heliospheric, and astrospheric.
  In the three component fits, the temperature and column density
  of the heliospheric absorption ($T=38,000\pm 8000$ K and
  $\log N_{\rm H}=14.6\pm 0.3$) are consistent with the measurements made
  for the $\alpha$~Cen line of sight.
\end{description}

\acknowledgments

     We would like to thank the anonymous referee for several helpful
suggestions.  Support for this work was provided by NASA through grant number
GO-07262.01-99A from the Space Telescope Science Institute, which is operated
by AURA, Inc., under NASA contract NAS5-26555.

\clearpage

\begin{deluxetable}{lcccccc}
\tablecaption{Fit Parameters\tablenotemark{a}}
\tablecolumns{7}
\tablewidth{0pt}
\tablehead{
  \colhead{Ion} & \colhead{$\lambda_{rest}$\tablenotemark{b}} &
    \colhead{Source of} & \colhead{$v$\tablenotemark{c}} & \colhead{$b$} &
    \colhead{log N} & \colhead{$\chi^{2}_{\nu}$} \\
  \colhead{} & \colhead{(\AA)} & \colhead{Absorption} &
     \colhead{(km~s$^{-1}$)} & \colhead{(km~s$^{-1}$)} & \colhead{} &
     \colhead{}}
\startdata
\sidehead{One Component Fits}
Fe II & 2600.173 & LISM        & $-28.4\pm 0.4$ & $2.12\pm 0.61$ &
  $12.65\pm 0.25$ & 0.41 \\
Mg II\tablenotemark{d} & 2796.352 & LISM        & $-27.9\pm 0.2$ &
  $3.15\pm 0.25$ & $13.05\pm 0.15$ & 2.19 \\
Mg II\tablenotemark{d} & 2803.531 & LISM        & $-27.9\pm 0.2$ &
  $3.15\pm 0.25$ & $13.05\pm 0.15$ & 2.19 \\
D I   & 1215.339 & LISM        & $-29.0\pm 0.2$ & $7.33\pm 0.17$ &
  $13.00\pm 0.01$ & 1.12 \\
H I   & 1215.670 & LISM        & $-25.9\pm 0.1$ &$14.34\pm 0.02$ &
  $18.178\pm 0.002$& 2.30 \\
\sidehead{Two Component Fits}
H I   & 1215.670 & LISM        &   ($-28.4$)    &   (10.11)      &
  ($17.85\pm 0.15$)& 1.29 \\
      &          & Heliosphere & $-26.7\pm 0.3$ & $28.5\pm 2.5$  &
  $15.08\pm 0.15$ & 1.29 \\
\sidehead{Three Component Fits}
H I   & 1215.670 & LISM        &   ($-28.4$)    &   (10.11)      &
  ($17.85\pm 0.15$)& 1.25 \\
      &          & Heliosphere &  ($-15\pm 5$)  & $25.0\pm 2.6$  &
  $14.6\pm 0.3$   & 1.25 \\
      &          & Astrosphere &  ($-35\pm 5$)  & $26.0\pm 3.0$  &
  $14.7\pm 0.3$   & 1.25 \\
\enddata
\tablenotetext{a}{Quantities in parentheses are fixed rather than derived
  (see text for details).}
\tablenotetext{b}{In vacuum.}
\tablenotetext{c}{Central velocity in a heliocentric rest frame.}
\tablenotetext{d}{The Mg~II lines were fitted simultaneously and forced to
  have the same fit parameters.}
\end{deluxetable}

\clearpage

\begin{figure}
\plotfiddle{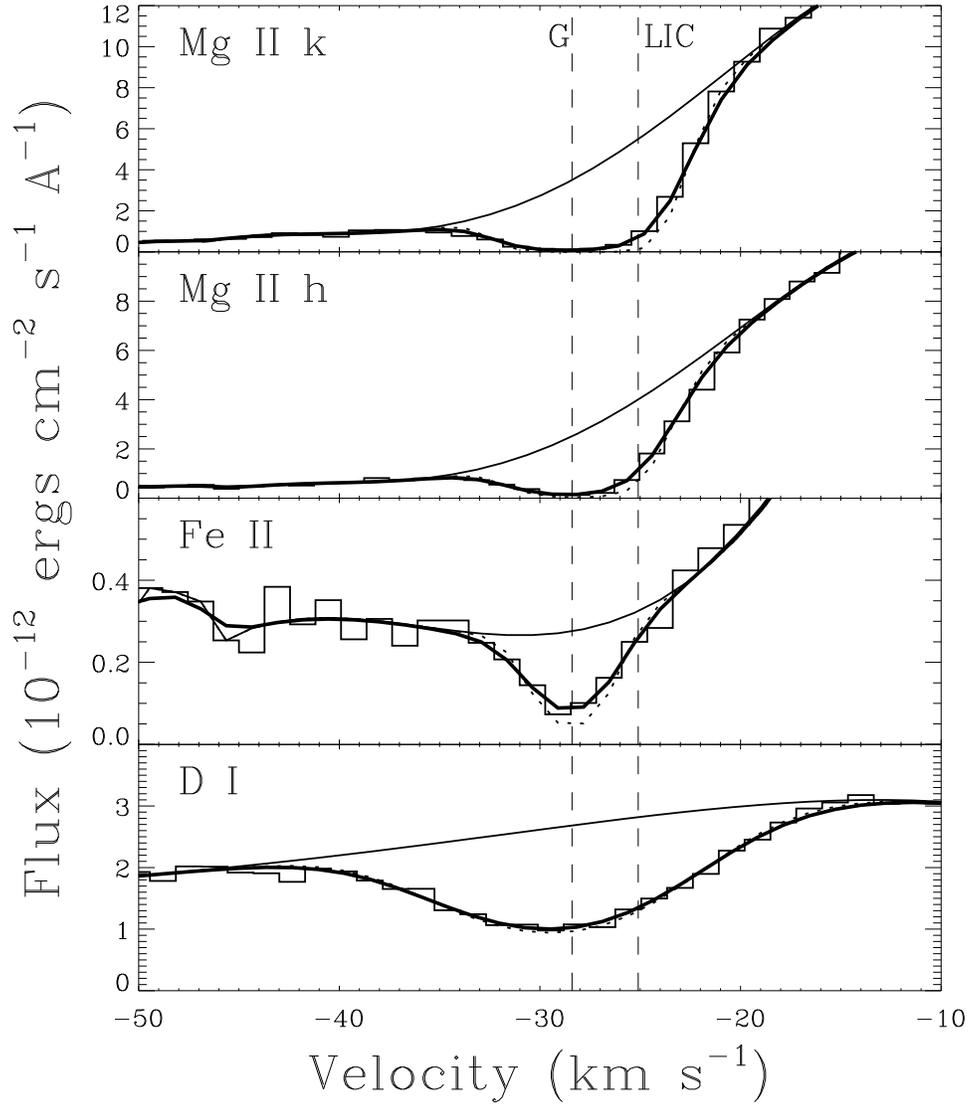}{6.0in}{0}{80}{80}{-290}{10}
\caption{Single component absorption line fits to the following lines:
  Mg~II k $\lambda$2796.352, Mg~II h $\lambda$2803.531, Fe~II
  $\lambda$2600.173, and D~I $\lambda$1215.339.  The dotted and thick solid
  lines are before and after instrumental broadening.  The data are displayed
  on a velocity scale in a heliocentric rest frame.  Dashed lines show the
  projected velocities of the LIC and G cloud flow vectors for the 36~Oph
  line of sight.}
\end{figure}

\clearpage

\begin{figure}
\plotone{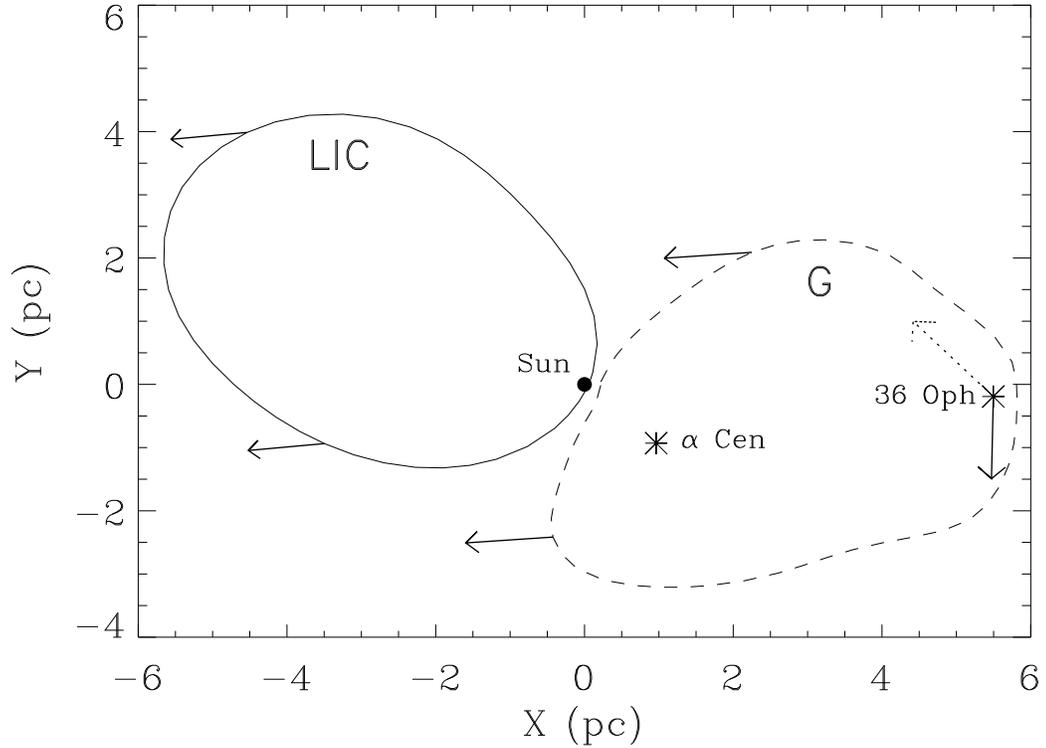}
\caption{A map showing the locations of the Sun, $\alpha$~Cen, 36~Oph, the
  LIC (solid line), and G cloud (dashed line), projected into the Galactic
  plane, where Galactic Center is to the right.  The LIC shape in the
  Galactic plane is from \citet{sr00}, while the G cloud shape is only a
  rough estimate.  The solid arrows indicate the velocity vectors of the LIC,
  the G cloud, and 36~Oph relative to the Sun, with the length of the arrow
  being proportional to the speed.  The dotted line indicates the
  G cloud vector in the rest frame of 36~Oph, thereby indicating the
  orientation of the star's astrosphere.  Note that the two stars and all
  the displayed vectors lie within $21^{\circ}$ of the Galactic plane, so
  this two-dimensional map is not greatly misrepresentative of the actual
  three-dimensional situation.}
\end{figure}

\clearpage

\begin{figure}
\plotfiddle{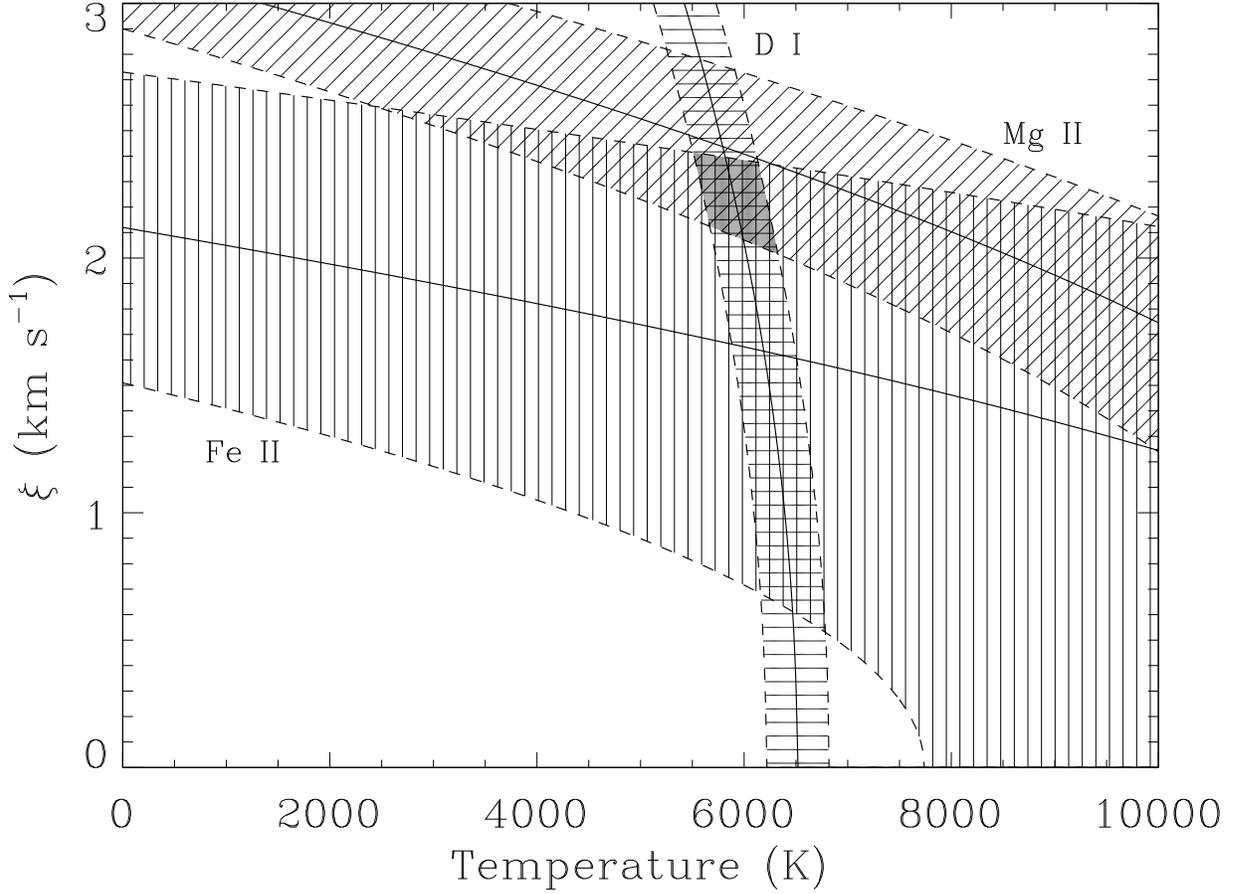}{3.5in}{90}{75}{75}{280}{0}
\caption{Nonthermal velocities ($\xi$) are plotted versus temperature,
  based on the Doppler parameters and their uncertainties measured from
  absorption lines of Fe~II, Mg~II, and D~I (see Table 1).  The shaded area
  where the three curves overlap indicates the actual temperature and $\xi$
  value of the interstellar material ($T=5900\pm 500$ K and
  $\xi=2.2\pm 0.2$ km~s$^{-1}$).}
\end{figure}

\clearpage

\begin{figure}
\plotone{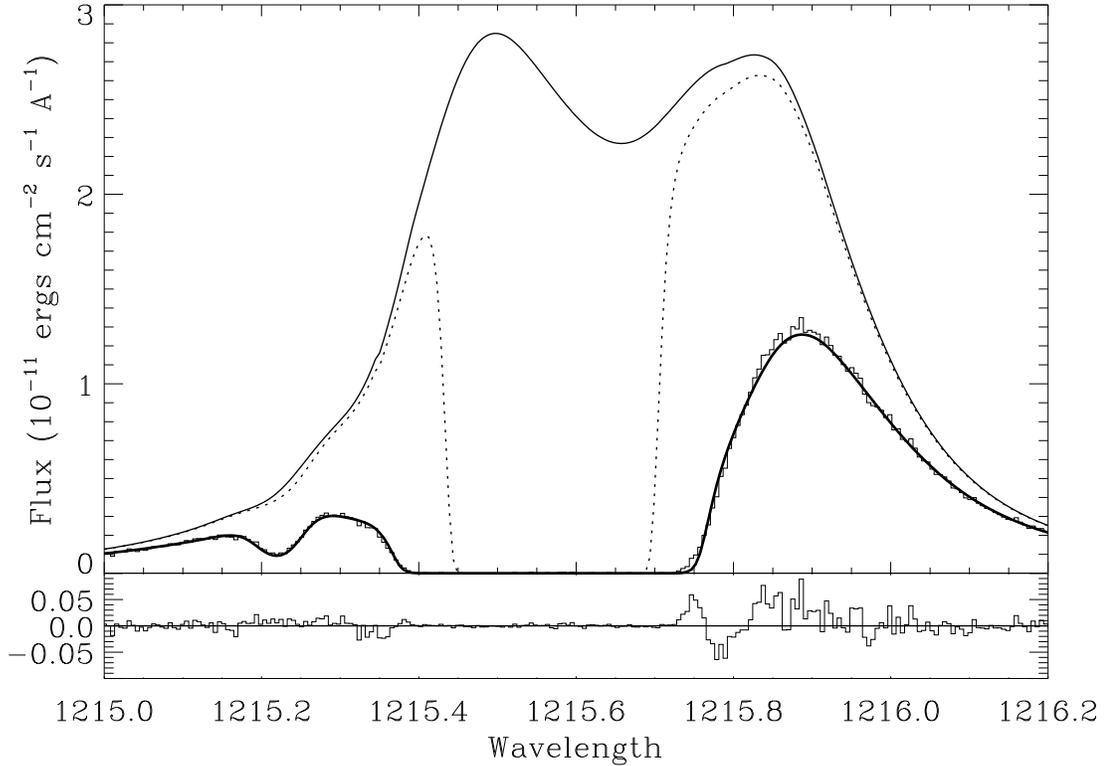}
\caption{A single component fit (thick solid line) to the broad H~I
  Lyman-$\alpha$ absorption line centered at 1215.6~\AA\ and the D~I
  Lyman-$\alpha$ absorption at 1215.2~\AA, which are shown on a
  heliocentric wavelength scale.  This fit is not realistic because
  the central velocity and temperature of the H~I line are inconsistent with
  D~I and the other LISM lines.  The dotted line shows the LIC absorption
  associated with the upper limit for the LIC column density derived in the
  text ($N_{\rm H~I}<6\times 10^{16}$ cm$^{-2}$).  Practically all of
  the absorption is well within the saturated core of the line, so LIC
  absorption will have no observable effect on the Lyman-$\alpha$ line.}
\end{figure}

\clearpage

\begin{figure}
\plotfiddle{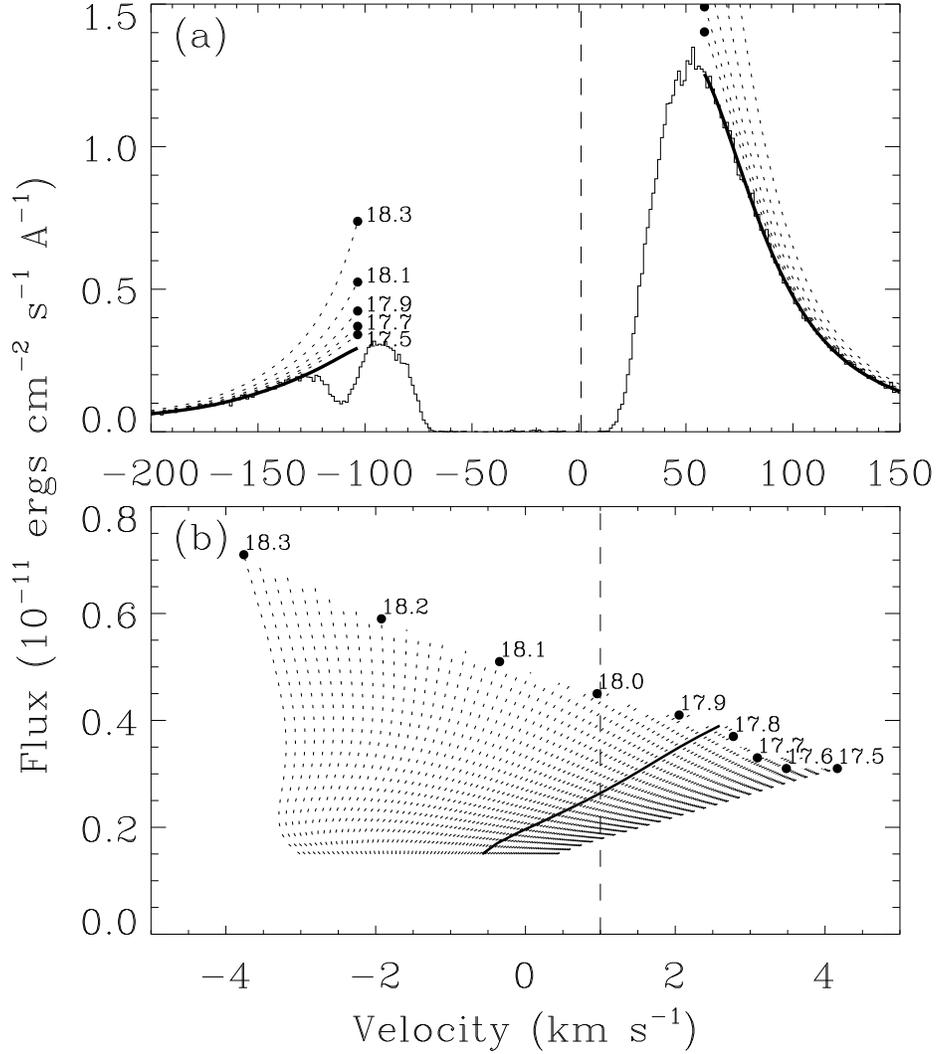}{4.5in}{0}{80}{80}{-270}{0}
\caption{In (a), the wings of the unabsorbed stellar Lyman-$\alpha$ line
  are derived by extrapolating upwards from polynomial fits to the observed
  profile (thick solid lines) assuming five different
  LISM H~I column densities in the range $\log N_{\rm H}=17.5-18.3$.  The
  bisectors of Lyman-$\alpha$ wings computed in this fashion are
  shown in (b), as a function of $\log N_{\rm H}$.  When the correct value of
  $\log N_{\rm H}$ is assumed, the bisector should be at the radial velocity
  of the star, which is shown as a dashed line in both panels.  The thick
  line in (b) indicates the bisector that best matches the radial velocity,
  which corresponds to $\log N_{\rm H}=17.84$.}
\end{figure}

\clearpage

\begin{figure}
\plotfiddle{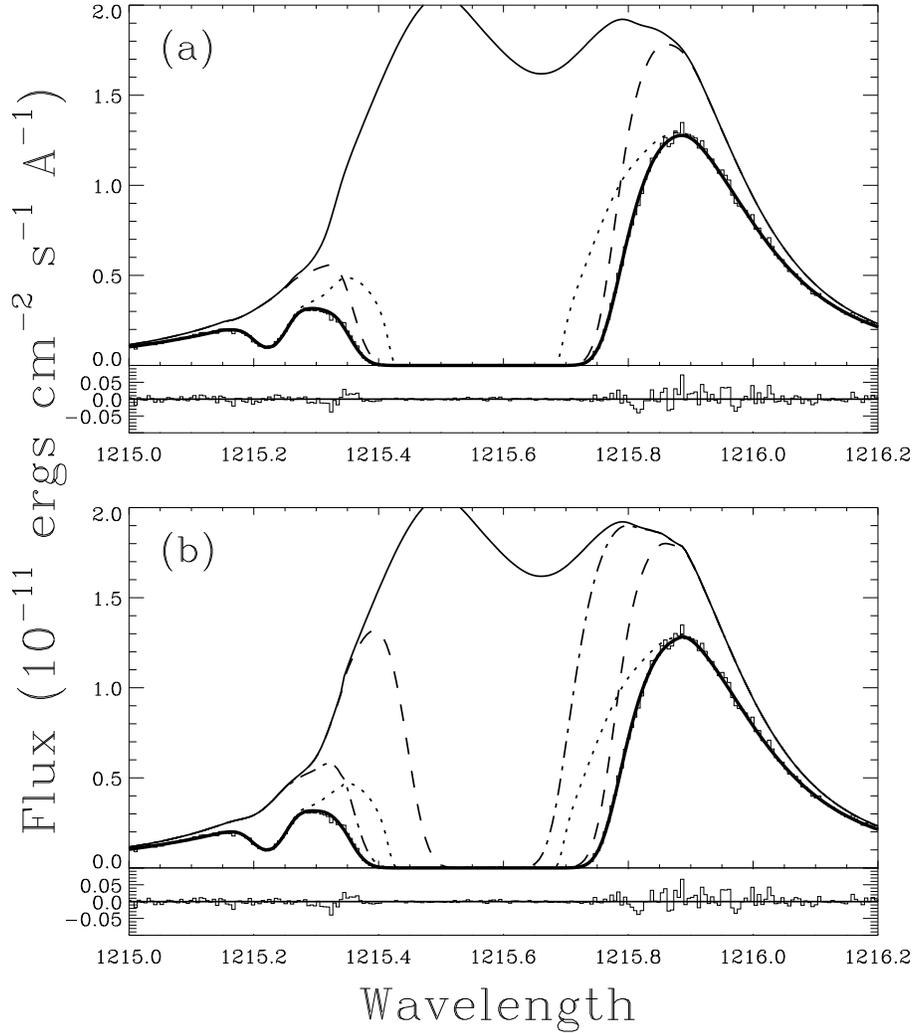}{4.5in}{0}{80}{80}{-270}{0}
\caption{(a) A two component fit to the H~I Lyman-$\alpha$ line, where
  the dotted line is LISM absorption and the dashed line is heliospheric
  hydrogen wall absorption.  The thick solid line is the combination
  of the two components, which fits the data.  Unlike the fit in Fig.\ 4, D~I
  and H~I are self-consistent in this fit.  The spectrum is displayed in a
  heliocentric rest frame.  (b) A fit similar to that in (a),
  but with the addition of a third component associated with absorption from
  hot astrospheric H~I surrounding 36~Oph (dot-dashed line).  We consider
  this to be our best model of the Lyman-$\alpha$ absorption (see text).}
\end{figure}

\end{document}